\documentstyle[emulateapj,apjfonts,psfig]{article}

\lefthead{J.~M.~Miller et al.}  
\righthead{Ionized Absorption in H~1743$-$322}

\received{Date}
\revised{Date}
\accepted{Date}

\journalid{vol}{date}
\articleid{1}{4}
\paperid{id}

\cpright{AAS}{1999}
\ccc{x}

\begin{document}

\title{Simultaneous {\it Chandra} and {\it RXTE} Spectroscopy of the
  Microquasar H~1743$-$322:\\ Clues to Disk Wind and Jet
  Formation from a Variable Ionized Outflow}

\author{J.~M.~Miller\altaffilmark{1,2},
        J.~Raymond\altaffilmark{1},
	J.~Homan\altaffilmark{3},
        A.~C.~Fabian\altaffilmark{4},
	D.~Steeghs\altaffilmark{1},
	R.~Wijnands\altaffilmark{5},
	M.~Rupen\altaffilmark{6},
	P.~Charles\altaffilmark{7},
	M.~van~der~Klis\altaffilmark{5},
	W.~H.~G.~Lewin\altaffilmark{3},	
	}

\altaffiltext{1}{Harvard-Smithsonian Center for Astrophysics, 60
	Garden Street, Cambridge, MA 02138, jmmiller@cfa.harvard.edu}
\altaffiltext{2}{NSF Astronomy and Astrophysics Fellow}
\altaffiltext{3}{Center~for~Space~Research and Department~of~Physics,
        Massachusetts~Institute~of~Technology, Cambridge, MA
        02139--4307}
\altaffiltext{4}{Institute of Astronomy, University of Cambridge,
        Madingley Road, Cambridge CB3 OHA, UK}
\altaffiltext{5}{Astronomical Institute ``Anton Pannekoek,''
        University of Amsterdam, and Center for High Energy
        Astrophysics, Kruislaan 403, 1098 SJ, Amsterdam, NL}
\altaffiltext{6}{National Radio Astronomy Observatory, Array
        Operations Center, 1003 Lopezville Road, Socorro, NM 87801}
\altaffiltext{7}{Department of Physics and Astronomy, University of
	Southampton, SO17 1BJ, England, UK}


\authoremail{jmmiller@cfa.harvard.edu}

\label{firstpage}

\begin{abstract}
We observed the bright phase of the 2003 outburst of the Galactic
black hole candidate H~1743$-$322 in X-rays simultaneously with {\it
Chandra} and {\it RXTE} on four occasions.  The {\it Chandra}/HETGS
spectra reveal narrow, variable (He-like) Fe XXV and (H-like) Fe XXVI
resonance absorption lines.  In the first observation, the Fe XXVI
line has a FWHM of $1800 \pm 400$ km/s and a blue-shift of $700 \pm
200$ km/s, suggesting that the highly ionized medium is an outflow.
Moreover, the Fe~XXV line is observed to vary significantly on a
timescale of a few hundred seconds in the first observation, which
corresponds to the Keplerian orbital period at approximately
$10^{4}~r_{g}$ (where $r_{g} = GM/c^{2}$).  Our models for the
absorption geometry suggest that a combination of geometric effects
and changing ionizing flux are required to account for the large
changes in line flux observed between observations, and that the
absorption likely occurs at a radius less than $10^{4}~r_{g}$ for a
$10~M_{\odot}$ black hole.  Viable models for the absorption geometry
include cyclic absorption due to an accretion disk
structure, absorption in a clumpy outflowing disk wind, or possibly a
combination of these two.  If the wind in H~1743$-$322 has 
unity filling factor, the highest implied mass outflow rate is 20\%
of the Eddington mass accretion rate.  This wind may be a hot
precursor to the Seyfert-like, outflowing ``warm absorber'' geometries
recently found in the Galactic black holes GX 339-4 and
XTE~J1650$-$500.  We discuss these findings in the context of ionized
Fe absorption lines found in the spectra of other Galactic sources,
and connections to warm absorbers, winds, and jets in other accreting
systems.
\end{abstract}

\section{Introduction}
The Galactic X-ray transient H~1743$-$322 was classified as a black
hole candidate by White \& Marshall (1984) based on its X-ray spectral
characteristics.  The source was discovered during its 1977--1978
outburst with {\it HEAO-1} and {\it Ariel V} (Doxsey et al.\ 1977,
Kaluzienski \& Holt 1977).  The next convincing detection of
H~1743$-$322 was made on 2003 March 21 with {\it INTEGRAL} (Revnivtsev et
al.\ 2003) and later with {\it RXTE} (Markwardt \& Swank 2003a),
earning the source the additional designations IGR~J17464$-$3213 and
XTE~J1746$-$322.  Markwardt \& Swank (2003b) pointed out that although
the original {\it HEAO-1} detection reported two possible locations,
it was very likely that the potentially new transient was actually
H~1743$-$322.  Following its 2003 re-activation in X-rays, the
source was extensively monitored in optical, IR, and radio bands (see,
e.g., Rupen, Mioduszewski, \& Dhawan 2003; Steeghs et al.\ 2003 and
ref. therein).  

Two discoveries made during the 2003 outburst of H~1743$-$322 have
provided additional evidence that this source harbors a black hole
primary.  First, Homan et al.\ (2003) reported the detection of
quasi-periodic oscillations (QPOs) at 240~Hz in the {\it RXTE} X-ray
flux.  This frequency is typical of black hole candidates, and is
tantalizingly close to the Keplerian orbital frequency expected at the
innermost stable circular orbit (ISCO) around a Schwarzschild black
hole ($\nu \simeq 220~Hz~(10~M_{\odot}/M_{BH})$).  Moreover, there is
evidence for QPOs at 160~Hz, and therefore in the 2:3 ratio which
has recently emerged in a few Galactic black hole systems
(GRO~J1655$-$40: Strohmayer 2001a, GRS~1915$+$105: Strohmayer 2001b,
XTE~J1550$-$564: Miller et al. 2001, and possibly XTE J1650$-$500:
Homan et al. 2003b).  The apparent 2:3 pairing may be evidence of a
parametric epicyclic resonance in black hole accretion disks (see,
e.g., Abramowicz \& Kluzniak 2003).  However, there is evidence that
black hole QPO pairs may drift out of a 2:3 ratio (Miller et al.\
2001, Remillard et al.\ 2002), and the 2:3 ratio described by the resonance
model for so-called kHz QPOs in neutron star systems is at odds with
data (Miller 2003).

Second, radio observations have revealed a jet in H~1743$-$322, moving
at $v/c \simeq 0.8$ (Rupen, Mioduszewski, \& Dhawan 2004).  On the
basis of this relativistic jet, H~1743$-$322 may be termed a
``microquasar''.  Indeed, the X-ray spectroscopic results presented in
this paper will further establish close similarities between
H~1743$-$322 and microquasars like GRO~J1655$-$40 and GRS~1915$+$105,
in addition to their radio jet and X-ray timing similarities.
Although highly relativistic jets are not a unique black hole primary
signature (Fender et al.\ 2004), such jets are are more commonly
observed in black hole systems than in neutron star systems.

We observed H~1743$-$322 simultaneously with {\it Chandra} and {\it
RXTE} on four occasions during the bright phase of its 2003 outburst,
as part of our larger multi-wavelength black hole observing program.
Herein, we report the results of an analysis of
the X-ray spectra resulting from these observations.  Most
importantly, the {\it Chandra}/HETGS spectra reveal variable He-like
Fe~XXV and H-like Fe~XXVI resonance absorption lines, which may be due
to absorption in an outflow.  Variable, Seyfert-like ``warm-absorber''
outflows have recently been detected in the X-ray spectra of the
Galactic black holes XTE~J1650$-$500 and GX~339$-$4 (Miller et al.\
2004a, Miller et al.\ 2004b).  We examine the absorption in
H~1743$-$322 within the context of these recent results.  The results
of an analysis of the X-ray timing properties of the {\it RXTE}
observations will be reported in a companion paper (Homan et al.\
2004).  The results of radio observations made in partial
support of this program will be reported in a separate paper by Rupen
et al. (2004, in prep.), and results from optical and IR observations
will be reported in Steeghs et al.\ (2004, in prep.).

\section{Observation and Data Reduction}
We observed H~1743$-$322 with {\it Chandra} and {\it RXTE} on four
occasions during the bright phase of its 2003 outburst, as part of
ongoing and complementary Target of Opportunity programs to study
Galactic black holes in outburst in X-rays.  The {\it Chandra}
observations were each approximately 50~ksec in duration.  {\it RXTE}
made observations simultaneous with each of these {\it Chandra}
observations, with individual exposure times ranging between 0.6~ksec
and 25~ksec.  The observation dates, starting times, and net exposure
times are given in Table 1.

\subsection{{\it RXTE} Spectral Reduction}
At the time of writing, PCU-2 is the best-calibrated PCU in the {\it
RXTE}/PCA, based on simple power-law fits to the Crab spectrum.  In
order to best search for subtle features like broad Fe~K$\alpha$ lines,
we restricted our PCA spectral reduction and analysis to PCU-2.
Spectra from the HEXTE-A cluster were included for spectral reduction
and analysis to constrain the high energy continuum.  {\it RXTE} data
reduction tools in the HEASOFT version 5.2 suite were used to screen
the event and spectral files.

Spectra from PCU-2 were extracted from data taken in ``Standard 2''
mode, providing full coverage of the 2.0-60.0~keV PCA bandpass in 129
channels every 16 seconds.  Data from all of the Xe gas layers in
PCU-2 was combined.  Background spectra were made using the tool
``pcabackest'' using the latest ``bright source'' background model.
Redistribution matrix files (rmfs) and ancillary response files (arfs)
were made and combined into a single file using the tool ``pcarsp.''
The background spectra were subtracted from the total spectrum using
the tool ``mathpha.''  It is well-known that fits to PCA spectra of
the Crab with a simple power-law model reveal residuals as large as
1\%, but adding 1\% systematic errors to PCA data is often an
over-estimate.  Using the tool ``grppha'', we added 0.6\% systematic
errors to the spectra from PCU-2.  The lowest energy bins in the PCA
are poorly calibrated, and the calibration above 25~keV is uncertain;
we therefore restricted our PCU-2 spectral analysis to the 3--25~keV
range.

The HEXTE-A spectra were made from standard ``archive'' mode data,
which has a nominal time resolution of 32 s and covers the 10--250~keV
range with 61 channels.  All spectra were background-subtracted and
deadtime corrected using the standard procedures.  The calibration of
the energy channels below 20~keV is poor, and we therefore restricted
our analysis of the HEXTE spectra to energies above 20~keV.  As some
of the spectra are soft and therefore dominated by background above
100~keV, we fixed this energy as an upper bound for consistency.
However, analysis of the spectra up to 250~keV was undertaken for
those spectra with sufficient signal.

The {\it RXTE} spectra were analyzed using XSPEC version 11.2.0
(Arnaud \& Dorman 2000).  All errors on fits to the {\it RXTE} spectra
are 90\% confidence errors.

\subsection{{\it Chandra} Spectral Reduction}
In each {\it Chandra} observation, the High Energy Transmission
Grating Spectrometer (HETGS) was used to disperse the incident flux
onto the ACIS-S CCD array.  The ACIS-S array was read-out in
``continuous-clocking'' mode (nominal frametime: 2.8 msec) to prevent
photon pile-up.  Information is collapsed into one dimension to
achieve the fast read-out in this mode; thus, our {\it Chandra}
observations were purely spectroscopic as imaging information was not
preserved.  A 100-column ``gray filter'' window was used around the
aimpoint on the ACIS-S3 chip, which acted to read-out only only one in
every ten zeroth-order events.  This step was taken to prevent
telemetry saturation and frame drop-out on the S3 chip.  We shifted
the location of the aimpoint from the nominal S3 aimpoint.  A
Y-coordinate offset of 0.33 arcmin was used to shift the wavelength at
which CCD chip gaps occur in order to ensure a simple response
function in the Fe K line region, and a SIM-Z offset of $-$7.5~mm was
used to push the aimpoint toward the CCD read-out to limit charge
transfer inefficiencies and to limit wear and tear on the prime
imaging area of the S3 chip.  

Spectra and lightcurves were extracted from the {\it Chandra} event
lists using the CIAO version 2.3 reduction and analysis suite.  The
``evt1'' file was filtered to accept only the standard event grades,
to accept only events from the nominal good time intervals, and to
exclude bad pixels.  The ``destreak'' tool was run to remove the
effects of ``streaking'' in the CCDs.  Although imaging information is
not preserved in the mode used for these observations, the CIAO HETGS
spectral extraction routines are robust for locating the observation
aimpoint, defining a wavelength grid, and extracting spectra.  Spectra
were extracted using the default settings for the tools
``tg\_resolve\_events'' and ``tgextract.''  We used the canned
redistribution matrix files (rmfs) to make ancillary response files
(arfs) via the ``fullgarf'' tool.  The first-order HEG spectra and arfs
were added to make a single spectrum and single arf using the tool
``add\_grating\_spectra''; the same addition was done with the
first-order MEG spectra.  In the analysis detailed below, we only
claim spectral features which are readily detected in the individual
dispersed spectra (prior to addition).

The {\it Chandra}/HETGS spectra were fit with ISIS version 1.1.3
(Houck \& Denicola 2000).  All errors on the absorption line
parameters measured from the HETGS spectra are $1\sigma$ errors.

In this paper, we have restricted our analysis to the first-order HEG
spectra.  Moreover, we fit only phenomenological power-law continua
and simple Gaussian line functions in a range spanning less than 1\AA.
The primary reasons for these restrictions are (1) the low-energy
portions of the MEG and HEG spectra do not reveal any evidence for
narrow emission or absorption lines, and (2) only the HEG spectra have
sufficient sensitivity through the Fe~K line range to allow meaningful
analysis and constraints.  

For the benefit of scientists examining these spectra in the future,
we note some additional reasons behind the above restrictions, since
they are related to nuances of the telescope and the observing mode
employed.  H~1743$-$322 had a flux roughly equivalent to or greater
than 1~Crab in the {\it RXTE}/ASM (1.5--12.0~keV) during the first
{\it Chandra} observations, {\it despite} the high column density
along the line of sight to this source.  When fit with the spectral
shapes that describe the {\it RXTE} spectra well, the {\it
Chandra}/HETGS spectra show positive residuals above approximately
8~keV, and also at energies below approximately 2~keV.  These effects
are inconsistent with photon pile-up, and inconsistent with the
effects expected due to contaminant build-up on the ACIS chips.
Rather, the positive residuals above $\sim$8~keV are likely due to
incident photons scattering off the instrument support structure and
falling into the high energy portion of the dispersed spectra (which
is closest to the zeroth order).  The nature of the positive residuals
below $\sim$2~keV is less certain.  If the standard CIAO tools
over-correct for charge transfer inefficiency at our offset, such a
correction might partially account for the positive residuals.

While these problems do not affect our ability to search for narrow
emission and absorption lines through the full HETGS bandpass, we were
unable to fit the full HETGS continuum with meaningful models, and we
do not report such fits in this work.  We note, however, that in the
2.5--7.5~keV range, the {\it Chandra}/HETGS spectra broadly agree with
the {\it RXTE} spectra.  The positive residuals above $\sim$8~keV made
it impossible to measure the depth of the (He-like) Fe~XXV and
(H-like) Fe XXVI K edges at 8.8~keV and 9.3~keV (respectively).
Balancing the optical depth of these K-shell absorption edges with the
optical depth of the Fe~XXV and Fe~XXVI absorption lines we detected
(see below) would have allowed tighter constraints.  Nevertheless, a
number of important constraints can be made using only the variable
Fe~XXV and Fe~XXVI absorption lines detected in these spectra, and we
report measurements of the line parameters, a model for the
absorption, and implications for the accretion inflow and outflow in
H~1743$-$322 in this work.

\section{Analysis and Results}

\subsection{Broad-Band Spectral Fits}
We made fits to each of the time-averaged {\it RXTE} spectra listed in
Table 1 with simple, phenomenological spectral models.  Initial fits
revealed that an equivalent neutral hydrogen column density of $N_H =
2.3 \times 10^{22}~{\rm cm}^{-2}$ was easily within the error bars of
each individual measurement.  Therefore, we fixed this value in all
subsequent fits.  This column is higher than that predicted from radio
measurements ($N_{H} \simeq 7 \times 10^{21}~{\rm cm}^{-2}$, Dickey \&
Lockman 1990); however, the radio estimates are limited by the coarse
spatial resolution of the survey observations.  Our spectral model
consisted of the multicolor disk blackbody (MCD; Mitsuda et al.\ 1984)
and power-law components.  As the name implies, the MCD model is
included to account for soft, thermal, blackbody-like emission thought
to arise from an optically-thick but geometrically-thin accretion
disk.  The power-law component is likely primarily the result of
Compton-upscattering of disk photons by a central, hot,
electron-dominated ``corona''.  This region may not be distinct from
the base of a jet; if it is distinct, processes like synchrotron
self-Compton emission from the base of a jet may also contribute to
the hard power-law component.

The results of spectral fits in the 3--100~keV band with this absorbed
MCD plus power-law are reported in Table 2, and an example spectral
fit is shown in Figure 1.  The spectral parameters obtained through
our fits are typical of Galactic black hole candidates in bright
phases.  In general, statistically acceptable fits are obtained
($\chi^{2}/\nu \simeq 1$, where $\nu$ is the number of degrees of
freedom in the fit) with this simple model.  (The fits would be
statistically improved if the Xe edge at approximately 4.8~keV was
better described in the detector response.)  The disk temperature
varies between $kT = 1.0-1.2$~keV in our observations, indicating that
the disk is likely important throughout the bright phase of the
outburst and close to the innermost stable circular orbit (ISCO).  The
broad-band spectral properties indicate that H~1743$-$322 was likely
in an extended ``very high'' or ``steep power-law'' state in
observations 1a, 1b, 1c, 2a, 2b, 2c.  Observation 3 occurred in a
phase intermediate between the ``high/soft'' an ``very high'' (``steep
power law'') state, but was likely more similar to the ``high/soft''
(or, ``thermal-dominant'') state.  Observations 4a, 4b, and 4c likely
occurred in the ``high/soft'' (or, ``thermal-dominant'') state.  The
states observed in H~1743$-$322 differ in subtle ways from canonical
black hole states, however, and the reader is referred to Homan et
al.\ 2004 (in prep.) for a more complete analysis and discussion of
the states observed.

The highest flux inferred in the 0.5--10.0~keV band is $7.8 \pm 0.8
\times 10^{-8}~{\rm erg}~{\rm cm}^{-2}~{\rm s}^{-1}$, which
corresponds to a luminosity of $L_{X} = 6.8\pm 0.7 \times
10^{38}~(d/8.5~{\rm kpc})^{2}~{\rm erg}~{\rm s}^{-1}$ (see Table 2).
At present, neither the mass of H~1743$-$322 nor the distance to the
source are known.  Given its Galactic center position and rather high
column density, however, it is reasonable to use a Galactic center
distance of 8.5~kpc as an estimate of the distance to the source.  At
such a distance, the source emission was (just) sub-Eddington for a
$6~M_{\odot}$ black hole (and more comfortably sub-Eddington for a
$10~M_{\odot}$ black hole).

Since Homan et al.\ (2003a) detected QPOs at 240~Hz and 160~Hz in
H~1743$-$322, which may be indicative of black hole spin as the QPOs
are in a 2:3 ratio (Abramowicz \& Kluzniak 2003), it is worth
addressing whether or not our spectral fits support the inference of a
disk extending very close to a potentially spinning black hole.  The
MCD model allows a measure of the inner disk radius through its
normalization: $K = [(r_{in}/{\rm km})/(d/10~{\rm kpc})]^{2} \times
cos(\theta)$~ (where $\theta$ is the inclination of the disk to the
line of sight).  The smallest disk normalization is measured in
observation 2b.  Via the normalization, $r_{in} \simeq 30$~km for
$d=8.5$~kpc and $\theta = 45^{\circ}$, and $r_{in} \simeq 50$~km for
$d=8.5$~kpc and $\theta=80^{\circ}$.  The ISCO around a non-spinning
black hole is $r_{ISCO} = 6~r_{g}$ (where $r_{g} = GM_{BH}/c^{2}$;
$r_{g} \simeq 1.5~{\rm km}~(M_{BH}/M_{\odot})$).  For black hole
masses greater than $10~M_{\odot}$, $R_{in} \leq 90$~km may imply
spin and support any such implications from the QPOs previously
detected.  However, this simple analysis has a few important problems
and the inferred radii should be regarded cautiously.  The MCD model
does not include an inner boundary condition, and inclination effects
and radiative processes in the disk can be important and have not been
accounted for in this exercise (see, e.g., Shimura \& Takahara 1995,
and Merloni, Fabian, \& Ross 2000).  Moreover, disk continuum
spectroscopy is particularly sensitive to the absorption model and to
the model for the hard component.  Finally, QPO frequencies may not be
simple Keplerian frequencies.

The broad-band {\it RXTE} spectra are remarkable for what they {\it do
not} show:

First, although we have only fitted the data up to 100~keV (since the
softer spectra have little flux above this energy), there is no
evidence for a break in the power-law component in any of the spectra.
Breaks may indicate the electron temperature of the putative corona.
The absence of such an observed break in these spectra may indicate
that the coronal electron temperature is generally very high ($kT
\simeq 100$~keV, or greater).  Observation 2b provides an illustrative
case.  This observed spectrum is reasonably hard and has a high S/N,
with H~1743$-$322 clearly detected out to 250~keV without a break.
Replacing the power-law component in our spectral model with the
compTT Comptonization model (in which the coronal electron temperature
is a variable, Titarchuk 1994), $kT_{e} \simeq 160$~keV is obtained.
The model underestimates the hard flux above 100~keV and is
statistically poor ($\chi^{2}/\nu = 136.4/76$), but indicates that the
coronal temperature may be very hot indeed.

Second, the spectra are free of any features expected to arise from
disk reflection (see Fig.\ 1).  Hard X-ray irradiation of the
(relatively) cool accretion disk by the corona is expected to produce
a characteristic set of features: an Fe~K$\alpha$ emission line, a
corresponding Fe~K edge, and a Compton-backscattering hump peaking
between 20--30~keV.  Disk reflection features are an extremely common,
if not ubiquitous feature of broad-band Galactic black hole X-ray
spectra.  Yet, there is no evidence for any of these features in any
of the broad-band {\it RXTE} spectra we obtained.  In Table 2, the
95\% confidence upper limits on the strength of narrow and broad
Fe~K$\alpha$ emission lines are given (FWHM$=0$~keV and FWHM$=2$~keV,
as per the Doppler broadening expected at the inner disk around a
Schwarzschild black hole).  In most cases, the upper limits are
stringent ($W_{K\alpha} \leq 15$~eV).  For a disk at the ISCO,
reflection from a neutral disk is expected to produce an emission line
with an equivalent width of $\sim 180$~eV (George \& Fabian 1991).
Especially in bright, high accretion rate phases, the disks around
Galactic black holes are likely to be ionized, and ionized disk
reflection models predict line equivalent widths that are a few times
higher than the neutral case (see, e.g., Ross, Fabian, \& Young 1999).

We made broad-band reflection fits to observation 2b to further
address the nature of the inner accretion flow geometry in
H~1743$-$322.  We replaced the simple power-law model with the
``constant density ionized disk model'' (CDID Ross, Fabian, \& Young
1999; Ballantyne, Ross, \& Fabian 2001).  The CDID model is
particularly well-suited to highly ionized reflection, and includes
Fe~K$\alpha$ line emission self-consistently.  We ``smeared'' the CDID
model by convolving it with the Laor relativistic emission line
function (Laor 1991).  The disk emissivity ($J(r) \propto r^{-q}$;
$q=3$ is expected for a standard disk), the inner disk radius $r_{in}$
(in units of $r_{g} = GM/c^{2}$), and inclination parameters in the
smearing function variables in the spectral model.  In all cases where
the reflection fraction $R = \Omega/2\pi$ was allowed to vary, values
along these lines were measured: log$(\xi) = 5.0-5.5$, $r_{in} \simeq
2~r_{g}$, $q \simeq 3$, $\theta = 70-85^{\circ}$, and $R = \Omega/2\pi
\leq 0.02$ (95\% confidence).  Such fits were formally acceptable
($\chi^{2}/\nu \simeq 1.0$) though clearly not statistically required
over the phenomenological MCD plus power-law model.  In fits with $R =
0.25$ fixed, even for $q=7$ (extreme smearing) the fit was not
acceptable $\chi^{2}/\nu \geq 2.0$.  Fits with higher values of $R$
fixed were statistically much worse.

\subsection{{\it RXTE} Variability Analysis}
The {\it RXTE} variability analysis conducted in this work follows the
methods and procedures detailed in Homan et al.\ (2004b).
The reader is directed to that work for a full discussion of the
timing analysis presented here.

An example lightcurve showing strong QPO-like variability on the
timescale of $few \times 100$~s is shown in Figure 2.  A power density
spectrum of {\it RXTE} observation 1a is shown in Figure 3; the
variability seen in the lightcurve corresponds to the broad $4
\times 10^{-3}$~Hz peak in the power density spectrum.  It is clear
that much of the power is concentrated in this feature.  The term
``QPO'' is usually reserverd for features with $Q\geq 2$ when fit with
a Lorentzian (where $\nu_{max} = \nu_{0}(1 + 1/4Q^{2})^{1/2}$).  As
$Q\simeq 0.5$ for the $4\times 10^{-3}$~Hz feature, it may not be
termed a QPO.  However, it is important to note that, like a QPO, its
frational rms {\it increases} with increasing energy (Homan et al.\
2004b; see Figures 2 and 5).

\subsection{{\it Chandra}/HETGS Spectral Analysis}
We began by analyzing the combined first-order HEG and MEG
time-averaged spectra from each observation in narrow 2\AA~ slices,
through the 1.2--25\AA~ range.  The continuum in each slice was fitted
with a simple phenomenological power-law model.  Where required,
absorption edge(s) due to neutral (atomic) photoelectric absorption in
the ISM were included and scaled to $N_{H} = 2.3\times 10^{22}~{\rm
cm}^{-2}$ (see Table 2).  Line features were fit with simple Gaussian
models.  Line identifications are based on the calculations and line
lists tabulated by Verner, Verner, \& Ferland (1996).  Column density
estimates were derived using the relation:

\begin{center}
$W_{\lambda} = \frac{\pi e^{2}}{m_{e} c^2} N_{j} \lambda^{2} f_{ij} =
8.85 \times 10^{-13} N_{j} \lambda^{2} f_{ij}$\\
\end{center}

\noindent where $N_{j}$ is the column density of a given species,
$f_{ij}$ is the oscillator strength, $W_{\lambda}$ is the equivalent
width of the line in cm units, and $\lambda$ is the wavelength of the
line in cm units (Spitzer 1978).  In using this relation, we have
assumed that the lines are optically thin and that we are on the
linear part of the curve of growth.

In observations 1, 3, and 4, we clearly detect absorption lines which
we identify as He-like Fe~XXV $1s^{2} - 1s2p$ ($\lambda = 1.850$\AA)
and H-like Fe~XXVI $1s-2p$ ($\lambda = 1.780$\AA) resonance absorption
lines.  The line wavelengths, velocity shifts, line breadths, fluxes,
equivalent widths, columns, and equivalent neutral hydrogen columns
are listed in Table 3.  The spectra are shown in Figure 4.  There is
no clear evidence for any narrow emission or absorption lines in the
rest of the spectrum.  There are some line-like features in the region
of the neutral Si edge which might be He-like Si XIII and H-like Si
XIV; however, these are more likely due to uncertainties in the
instrument response in this region (the spectrum peaks in this region,
and the CCDs contain Si).

We have measured significant variability in the line equivalent widths
and other parameters between observations, which demands changing
absorption and clearly ties the lines to a region close to (or within)
H~1743$-$322.  In observation 2, when the source flux was high and
twice as spectrally hard as in the other observations, the Fe~XXV and
Fe~XXVI lines are absent.  In observation 1, the lines are
significantly blue-shifted, suggesting that the absorbing matter was
flowing into our line of sight.  The Fe XXVI line in observation 3 is
also slightly blue-shifted.  In all other cases, the lines are
consistent with no net shift.  In observations 1 and 4, the Fe XXVI
lines are resolved, and have velocity widths (FWHM) of $1800\pm
400$~km/s and $1900\pm 500$~km/s, respectively.  The rest of the lines
are not resolved and likely have smaller velocity widths.

We made lightcurves of the dispersed spectrum from each observation in
the 0.5--10.0~keV band.  The lightcurves are presented in Figure 5.
Strong variability is present in all observations on the timescale of
$few \times 100$~s, but particularly in the first and second
observations.  This variability is confirmed to also be present in the
lightcurves of the simultaneous {\it RXTE} lightcurves of H~1743$-$322
(see Fig.\ 2 and Fig.\ 3).  A 10--12\% ``dip'' feature may be present
in the third {\it Chandra} observation.  A preliminary examination of
other {\it RXTE} observations in the public archive revealed much
stronger dipping activity typical of dipping black hole binaries
viewed at high inclinations (see, e.g., Kuulkers et al.\ 1998).

To explore whether the absorption lines vary within the dip, we made
spectra of observation 3 from before the dip and within the dip (see
Figure 5), and fit the spectrum in the manner noted above.  Again,
the only lines apparent are the Fe~XXV and Fe~XXVI absorption lines
found in the time-averaged spectrum.  The results of fits to the
spectrum prior to the dip and within the dip are given in Table 4,
and the spectra are shown in Figure 6.  Both absorption lines are
stronger within the dip.  While the statistical significance of the
variability in the Fe XXVI line is marginal, the variability in the
strength of the Fe XXV line is clearly significant.  Indeed, in
observation 3, the Fe XXV line is not clearly detected in the pre-dip
spectrum.

Next, we investigated whether the absorption lines in observation 1,
and 4 vary on the $few \times 100$~s flaring variability timescale.  We
calculated the mean count rate for observations 1, 2, and 4 (see
Fig.\ 5), produced lists of time intervals for count rates above and
below the mean rate, and extracted spectra from the time intervals
above and below the mean rates.  Those spectra were also fitted in the
manner described above; the spectra are shown in Figure 6.  As was the
case with the time-averaged spectrum of observation 2, the count-rate
selected spectra of observation 2 show no absorption lines.  In
observation 4, the strength of the lines does not vary significantly
between the high rate and low rate spectra.  In observation 1,
however, the absorption lines are marginally stronger in the low rate
spectra than in the high rate spectra.   This raises the interesting
possibility that the absorption may vary on a timescale of
$few \times 100$~s.  To establish this more clearly, we created a list of
time intervals in which the source count rate was more than 5 counts/s
above the mean rate, and more than 5 counts/s below the mean rate, and
extracted spectra from these time intervals.  The resultant spectra
are shown in Figure 7.  The Fe~XXV absorption line is clearly stronger
in the low count rate spectrum than in the high count rate spectrum,
and the variation is statistically significant (see Table 4).

\subsection{Photoionized Plasma Models}
To interpret the iron absorption lines measured in the time-averaged
spectra in more detail, we computed line profiles from a spherical
wind with a photoionization code.  We used the atomic physics packages
from the X-ray illuminated accretion disk models of Raymond (1993).
In this case, since the gas is optically thin, it is only necessary to
consider a single slab.  We used the spectral parameters taken from
Table 5 for each of the 4 observations and specify a density, slab
thickness and distance from the central object, to model a 1-d,
optically-thin wind.  The code iterates to find a self-consistent
temperature and ionization state, then computes equivalent widths
assuming a Doppler profile with widths ranging from 100 to 2000 $\rm
km~s^{-1}$.  It assumes ionization equilibrium, which is a good
approximation for the densities and dynamical times estimated below in
all cases.  From the output of the code we find the parameters that
match the observed Fe XXV and Fe XXVI absorption line equivalent
widths.
 
The Fe XXV and Fe XXVI lines cannot originate in exactly the same gas,
because the line widths and line shifts differ.  A plausible physical
picture might consist of denser, slower clumps containing more Fe XXV
embedded in a less dense gas where Fe XXVI dominates.  However, a
modest density contrast of order 2 (approximately) is sufficient to
shift the balance between Fe XXV and Fe XXVI, so an average density
that produces both lines is meaningful, and we do not have enough
measured parameters to justify the additional free parameters of a
more complex model (the line widths are uncertain).  We
therefore choose densities to match the observed equivalent widths and
Doppler velocity widths intermediate between those of Fe~XXV and
Fe~XXVI.
 
There is a maximum radius at which a slab can plausibly be causing the
observed absorption.  The Fe~XXVI/XXV line flux ratio gives an
ionization parameter ($\xi = L_{X} / nr^{2}$), and for any given
distance from the central source $r$, a specific density $n$ is
required.  A specific slab thickness, which cannot exceed $r$, is
required to produce the measured line column densities, so that $N
\leq nr$.  Because the density that gives the ionization parameter
scales as $r^{-2}$, the slab thickness scales as $r^{-2}$ (therefore
$N~r^{-1}$), and there is a maximum radius at which a slab of
thickness $r$ can reproduce the lines observed.  We call this
parameter $r_{max}$.  The lines can be formed at smaller $r$, and if
the lines are formed in a wind with $r^{-2}$ density profile, the
density at $r_{max}$ is correspondingly smaller.
 
Table 5 shows the derived parameters.  We can also compute the mass
loss rate in a wind of any velocity, because $n r^{-2}$ is constant.
The velocity shifts in Table 3 are generally uncertain, so we present
mass loss rates for a reference wind speed of 300 $\rm km~s^{-1}$.
 
The table shows that the lack of absorption lines in the second
observation cannot entirely be attributed to a higher ionizing flux.
Although the ionizing flux in this observation is about double that of
the first observation, an order of magnitude lower column density of
absorbing material is needed to match the upper limits to the
equivalent widths.
 
An interesting result in Table 5 is the small value of $r_{max}$
for the 4th observation, which results primarily from the small
ionizing flux.  The lines do not show a significant Doppler shift, but
they arise in a white dwarf-size region near the central source.  In
the first and third observations, the absorbing regions have a size
comparable to a few solar radii, which is similar to the separation of
the components in several binary black hole systems.
 
The second interesting aspect of the table is the indication that the
observed absorption lines are formed in a wind with a mass loss rate
of order $2 \times 10^{-8}~M_{\odot}~{\rm yr}^{-1}$ --- comparable to
the inferred mass accretion rate.  If we assume a distance of 8.5~kpc,
take the highest inferred 0.5--10.0~keV luminosity of $L_{X} = 6.8
\times 10^{38}~{\rm erg}~{\rm s}^{-1}$ and assume an efficiency of
10\% in the equation $L_{X} = \eta \dot{m}_{acc} c^{2}$, we obtain
$\dot{m}_{acc} = 7.6 \times 10^{18}~{g}~{\rm s}^{-1}$.  Taking the
highest measured 0.5--10.0~keV luminosity to be a lower limit on the
true Eddington luminosity, the highest inferred mass loss in the wind
($\dot{m}_{wind} = 1.7 \times 10^{18}~{\rm g}~{\rm s}^{-1}$)
corresponds to an outflow rate that is 22\% of the Eddington mass
accretion rate (for a unity filling factor).  Note, however, that the
fraction depends crucially on the filling factor, the energy range
used, and the distance assumed.  An outflow of this kind may be
analogous to the accretion disk winds from cataclysmic variables in
their high states and to T Tauri stars in high states (FU Ori stars),
which have wind mass loss rates of order 10\% of their accretion
rates.  However, Observation 4 shows a much smaller mass loss rate
along with its lower luminosity.

\section{Discussion}
\subsection{Geometric Constraints from the Broad-Band Spectral Fits}
The high equivalent neutral hydrogen column density along the line of
sight is consistent with a central Galactic distance to H~1743$-$322.
For distances greater than $\simeq$3~kpc, the highest 0.5--10.0~keV
unabsorbed flux observed (see Table 2) exceeds the Eddington limit for
a neutron star.  Moreover, the parameters obtained from broad-band
spectral fits to H~1743$-$322 with the phenomenological MCD plus
power-law model are typical for Galactic black hole systems in bright
phases (for a review, see McClintock \& Remillard 2003, see also Homan
et al.\ 2001).  These facts --- and the presence of 240~Hz QPOs in
H~1743$-$322 --- make it a near certainty that H~1743$-$322 harbors a
black hole primary.  However, the apparent absence of reflection
features in the spectrum, even in phases with 1) a strong hard
component, 2) a small inferred inner disk radius, and 3)
high-frequency QPOs, is at odds with the majority of Galactic black
holes.

It is likely that the low reflection fraction obtained is due to a
combination of astrophysical effects (ionization and disk inclination)
and the limitations of our reflection modeling.  Zdziarski et al.\
(2003) have shown that when the disk is fully ionized, the reflected
spectrum is nearly featureless, and might therefore be confused with a
zero--reflection scenario.  High disk ionization tends to smooth-out
reflection features; the dips we have detected in the lightcurve of
{\it Chandra} observation 3 and in additional public {\it RXTE}
observations of H~1743$-$322 point to a system viewed at high
inclination. At high inclinations, Doppler shifts become very
important, and tend to smear-out reflection features.  The CDID
reflection model we fit to {\it RXTE} observation 2b indicates that
the disk in H~1743$-$322 is likely highly ionized and seen at a high
inclination.  While these constraints are likely robust, it must be
acknowledged that even the CDID model has limitations.  It is
implemented into XSPEC as a library of angle--averaged spectra.
Finally, we note that if any reflection spectrum also passes through
the hard X-ray emitting corona, which is likely if a corona hugs a
disc viewed at high inclination, then the reflection component will be
Comptonized and much less detectable (Petrucci et al 2001).  It is
possible, then, that the total reflection model --- the CDID model
smeared with a line function (for which inclination is a fit
parameter) --- does not address high inclinations correctly.  Were the
combination of ionization and inclination effects more accurately
described by our model, it is likely that a high reflection fraction
would have been within errors.

The best interpretation of our broad-band spectral fits, then, is that
H~1743$-$322 is a black hole system seen at high inclination, with a
very highly or completely ionized inner disk.

\subsection{On the Nature of the Ionized Absorber}
We have clearly detected variable, narrow Fe~XXV and Fe~XXVI resonance
absorption lines in the spectra of H~1743$-$322.  In the recent past,
similar absorption lines have been detected in the microquasars
GRO~J1655$-$40 (Ueda et al.\ 1998), and GRS~1915$+$105 (Kotani et al.\
2000) with {\it ASCA}, and in GRS~1915$+$105 with {\it Chandra} (Lee
et al.\ 2002).  Unfortunately, each of these prior spectra suffered
heavily from photon pile-up.  The extraordinary sensitivity of the
H~1743$-$322 spectra has allowed us to improve upon previous
implications based on Fe XXV/XXVI absorption lines in microquasars in
two important ways: First, we have measured blue-shifts in the lines
detected in observation 1, providing direct evidence that the highly
ionized absorbing geometry may be an outflow (previously, outflows were
only asserted).  Second, we have resolved the lines in two cases, and
in other cases the lines are likely narrower than the instrumental
resolution.  The FWHM of the lines is vital for constraining the
nature of the absorption.

Viable descriptions of the absorption geometry in H~1743$-$322 must be
able to account for a number of our key findings, including: (1) the
blue-shifts observed in the first {\it Chandra} observation, (2) the
large line flux changes observed between observations (in particular,
the absence of absorption lines in the second {\it Chandra}
observation), and the stronger absorption seen in the dip observed
during the third {\it Chandra} observation, and (3) the absence of Fe
emission lines in the spectra, (4) the variability of the absorption
lines in observation 1 on a timescale of a few hundred seconds, and
(5) the absorption must occur at a radius smaller than the maximum
radius at which our photoionization models suggest the lines may be
produced.  Two geometries satisfy or
partially satisfy most of these requirements: cyclic absorption in a
disk structure resulting from the high inclination, and absorption in
an (potentially clumpy) outflowing disk wind.

The absence of Fe emission lines in the {\it Chandra} and {\it RXTE}
spectra suggests that the absorbing geometry is not spherical (else,
emission line photons would have scattered into our line of sight).
Indeed, modeling of the absorbers in GRO~J1655$-$40 (Ueda et al.\
2004) and GRS~1915$+$105 (Kotani et al.\ 2000) also suggested a
cylindrical, perhaps pancake-like absorbing geometry.  Absorption in a
disk structure --- perhaps a disk atmosphere --- is consistent with
this constraint.  However, this picture relies on relatively large
covering and filling factors.  The lack of emission lines in the
spectra is also consistent with the absorption occurring in relatively
dense clumps in a disk wind, with large covering factors but small
filling factors.  A clumpy wind is particularly appealing when one
considers that at high mass accetion rates (such as in the first {\it
Chandra} observation), our models for the absorber indicate that the
wind would already be optically-thick at approximately $300~r_{g}$.
Yet, we clearly see continuum spectral components consistent with
viewing the innermost flow without significant obscuration.  Moreover,
if QPOs are tied to Keplerian frequencies, the QPOs detected (Homan et
al.\ 2004, in prep.) indicate that emission from the innermost flow is
either seen directly or scattered into our line of sight.  A
significant fraction of the absorption needs to occur in clumps with a
high covering factor but small filling factor to be consistent with
the observed continuum spectral properties and variability properties.

The absence of absorption lines due to a combination of geometric
effects and ionizing flux (as implied by our photoionization models)
in the second {\it Chandra} observation is easily explained by the
clumpy outflow model.  Simply, a clump may not have intersected our
line of sight to the central accretion observation within this
observation.  It is more difficult to explain the absence of
absorption lines if the absorption occurs via cyclic
absorption in a disk structure.  The disk structure would have to be
dampened or disappear in the second {\it Chandra} observation.
However, the ionizing flux is higher in this observation, and
it has been shown that irradiation can induce warps in accretion disks
(see, e.g., Maloney, Begelman, \& Pringle 1996).  

To explain the enhanced absorption line fluxes measured
within the dip in the third {\it Chandra} observation, it is easy to
envision a cool clump absorbing Fe~XXV more strongly than the hotter
surrounding wind.  Alternatively, it is also possible that the dip
observed is due to an unrelated geometric effect, and that the
enhanced absorption line flux seen within the dip occurs because the
wind along our line of sight is temporarily shielded and allowed to
cool.  If the absorption lines are instead due to absorption in a disk
structure, a second warp or disk structure may need to be invoked to
explain the dip itself.  

The variability of the Fe XXV absorption line flux on timescales of a
few hundred seconds in the first {\it Chandra} observation can be
explained by both models.  Indeed, this short timescale line
variability is one of the most compelling arguments in favor of cyclic
absorption in a disk structure.  The period of the line variability
broadly agrees with the Keplerian orbital period at $10^{4}~r_{g}$ for
a $10~M_{\odot}$ black hole.  If H~1743$-$322 harbors a black hole
with $M \simeq 10~M_{\odot}$, then $10^{4}~r_{g}$ corresponds to
$\simeq 1.5 \times 10^{10}$~cm, which is less than $r_{max}$ in
observations 1 and 3 (see Table 5).  However, the observed variability
cannot be termed a QPO, and it is therefore less easily tied to a
cyclic disk structure than if it qualified as a true QPO.  If the
absorption occurs in a disk structure, it would suggest that the
variability itself is due to changes in the absorption in our line of
sight.  This is at odds with the fact that similar variability is
observed in the absence of absorption lines in the second {\it
Chandra} observation.  If the variability were due to absorption, the
variability should be more pronounced at low energies; in fact, the
fractional rms of the peak at $4\times 10^{-3}$~Hz increases with
higher energy (Homan et al.\ 2004b; see Figures 2 and 5).
Although it is less exciting, absorption in a clumpy outflowing wind
can likely account for the rapid line flux variability at least as
well as cyclic absorption in a disk structure.  Simply, the gas
ionization and recombination timescale is much faster than the
variability timescale observed.  If the variability is not due to a
disk structure, but rather just due to fluctuations in the intensity
of the central engine, an absorbing clump may be able to rapidly
respond to the changing flux.

Finally, the blue-shifts observed in the first {\it Chandra}
observation are more easily explained through an outflowing disk wind,
than by absorption in the disk.  The modest blue-shifts of the lines
are a small fraction of the speed of light, and similar to the
blue-shifted lines observed in GX~339$-$4 and XTE~J1650$-$500 (Miller
et al.\ 2004a), which are consistent with a disk wind.  If the
absorption is instead due to a cyclic disk structure, it is rather hard
to explain the blue-shifts without significant fine-tuning.  If the
structure were a warp, its motion should be tangential to our line of
sight, not into our line of sight.  Instead, the absorption might need
to occur in a disk wind flowing outward along the disk, and further
postulate that the wind density is modulated by a cyclic disk
structure underlying the atmosphere.  It has been shown that accretion
disks may drive outflows with a significant component along the disk
(see, e.g., Murray \& Chiang 1997, Proga 2003).  

On balance, absorption in an outflowing, clumpy wind appears to be the
more compelling than absorption in a disk structure.  Both
possibilities require a degree of fine-tuning to satisfy the
observational constraints.  A combination of these simple models may
also be a viable description of the absorption geometry.
 
\subsection{Connections to Warm Absorber Geometries}

The highly ionized outflow revealed in these {\it Chandra}
observations of H~1743$-$322 is very likely connected to the variable,
outflowing, Seyfert-like warm absorbers recently discovered in GX
339$-$4 and XTE~J1650$-$500 (Miller et al.\ 2004a, Miller et al.\
2004b).  In those systems, the outflowing absorbers are consistent
with disk winds or shell ejections.  The absorbers in GX~339$-$4 and
XTE~J1650$-$500 are blue-shifted, and so clearly outflowing.  In
GX~339$-$4, He-like Ne IX, Mg XI, and O VII absorption lines were
detected simultaneously with more strongly blue-shifted Ne II and Ne
III lines, suggesting a ``clumpy'' outflow with regions of differing
temperature and density.  The Fe~XXV and Fe~XXVI lines in H~1743$-$322
also need to come from distinct regions, though the contrast needed is
much less than for the wind observed in GX~339$-$4.  The outflow
observed in GX~339$-$4 was likely detected at a lower fraction of the
Eddingtion mass accretion rate in that source; its lower ionization
can therefore be attributed to a lower central source flux.  Moreover,
the outflow in GX~339$-$4 implied an order of magnitude lower mass
outflow rate than than the highest rate implied in H~1743$-$322
(perhaps as high as 22\% of the Eddington inflow rate).  The outflow
in H~1743$-$322 may be clumpy, however, and have a low filling factor;
in that case, the outflow rate in H~1743$-$322 could be lower by a
factor of $10^{1}-10^{2}$, or more.

It is plausible that the outflow in H~1743$-$322 is a hot, high
$\dot{m}_{wind}/\dot{m}_{Edd.}$ precursor to the cooler, lower
$\dot{m}_{wind}/\dot{m}_{Edd.}$ outflows observed in GX~339$-$4 and
XTE~J1650$-$500 (Miller et al.\ 2004a, 2004b).  The mass outflow rate
in the fourth {\it Chandra} observation of H~1743$-$322 is far less
than that in the first observation (see Table 5), further indicating
an evolution along the lines we suggest here.  The outflows in these
black hole systems are likely disk winds, and resemble the warm
absorbers revealed in some Seyfert galaxies (see, e.g., Reynolds 1997,
Elvis 2000, Morales \& Fabian 2002).  Such absorption regions could be
tied to rather equatorial disk outflows (Murray \& Chiang 1997, Proga
2003), rather than to a highly collimated flow perpendicular to the
disk.  Focused studies of Galactic black hole winds in the future may
enable us to learn about the evolution Seyfert warm absorbers; their
evolution is difficult to study directly given the inherently longer
timescales.

\subsection{On Possible Connections Between Ionized Absorption and Jets}

It is worth addressing the possibility that a putative disk structure
and the outflow are not merely coincident, but potentially causally
related.  For instance, a family of models has been discussed wherein
jets or a wind can be launched vertically from a disk through a
coupling between a spiral density wave in the disk and a Rossby vortex
(Tagger \& Pellat 1999; Varniere, Rodriguez, \& Tagger 2002;
Rodriguez, Varniere, Tagger, \& Durouchoux 2002).  It is possible,
then, that the $few \times 100$s variations seen in H~1743$-$322 are
due to a density structure which acts to drive jets close to the black
hole, and that the same wave modulates a disk wind far from the black
hole.  It is not yet clear if such a mechanism could be at work in
driving both jets and a disk wind, but if so, then the appearance of
ionized Fe absorption lines in strong jet sources like GRO~J1655$-$40,
GRS~1915$+$105, and now H~1743$-$322 would be explained naturally.

We note that using reduction methods exactly like those detailed for
H~1743$-$322, we have detected Fe~XXV and Fe~XXVI lines in a 60~ksec
{\it Chandra}/HETGS spectrum of Cygnus X-1 obtained on 3 March 2004,
when the source was in a canonical low/hard state (Miller et al.\
2004c, in prep.).  Cygnus X-1 is known to produce a compact, steady
radio jet in this state (Stirling et al.\ 2001).  Highly ionized Fe
absorption lines have also been detected in the neutron star system GX
13$+$1 (Sidoli et al.\ 2002).  Interestingly, GX~13$+$1 is also a
radio emitter in which a delay between its X-ray spectral hardness and
radio brightness has been observed which is very similar to that seen
in GRS~1915$+$105 (Homan et al.\ 2004a), strongly suggesting a compact
jet is also at work in this system.

However, simultaneous radio observations of H~1743$-$322 with the
VLA/VLBA reveal that the radio source was quite strong during the
first and second {\it Chandra} observations --- nearly 10~mJy at GHz
frequencies) --- but only 2~mJy during the third observation, and less
than 1 mJy during the fourth observation (Rupen et al.\ 2004, in
prep.).  Thus, strong radio activity (indicative of a jet) is seen
during the second {\it Chandra} observation (where absorption lines
are not observed), and absorption lines are clearly observed in the
third and fourth {\it Chandra} observations when the radio flux was
low.  At minimum, this suggests that if jets and ionized absorption
are causally related, they are not related in a simple way.  Both may
merely be the effect of a strongly active corona, for instance.

Moreover, the detection of highly ionized Fe absorption lines in high
inclination neutron star binaries may signify that it is unlikely that
ionized absorption lines are tied to jet production through a
mechanism like AEI.  Similar lines have also been seen in the
persistent emission of the (edge-on) dipping neutron star binaries
X~1254$-$690 (Boirin \& Parmar 2003), X~1624$-$490 (Parmar et al.\
2002), MXB~1658$-$29 (Sidoli et al.\ 2001) and 4U~1916$-$053 (Boirin
et al.\ 2004).  In contrast to the absorption lines in the black hole
systems H~1743$-$322, GX~339$-$4, and XTE J1650$-$500, the absorption
lines in these edge-on neutron star binaries are not significantly
blue-shifted.  Although the appearance of lines in both black hole and
neutron star systems suggests a related physical origin, the lack of
blue-shifts in the lines seen in neutron stars may hint at a different
physical picture.  Without significant blue-shifts, it is difficult to
clearly tie the lines to an outflow.  The lines may not be due to
absorption in an outflow, but might be due to absorption in ambient
gas within the system.  The absorption lines in MXB~1658$-$29, for
instance, are not observed to vary with orbital phase or even during
dipping events (Sidoli et al.\ 2001), which suggests that the
absorbing gas is not as close to the compact object as the absorbing
gas in black hole systems.\\

We wish to thank CXC Director Harvey Tananbaum and the CXC staff, and
Jean Swank and the {\it RXTE} staff, for making these TOO
observations.  We thank John Houck for assistance with ISIS.  JMM
acknowledges helpful discussions with Mike Nowak, David Huenemoerder,
Aneta Siemiginowska, Rob Fender, Herman Marshall, and Patrick
Wojdowski.  JMM gratefully acknowledges support from the NSF through
its Astronomy and Astrophysics Postdoctoral Fellowship program.  WHGL
and JH gratefully acknowledge support from NASA.  DS acknowledges
support from the SAO Clay Fellowship.  This research has made use of
the data and resources obtained through the HEASARC on-line service,
provided by NASA-GSFC.

\clearpage

\begin{table}[t]
\caption{Observation Log}
\begin{footnotesize}
\begin{center}
\begin{tabular}{lll}
Instrument & Start Time & Net Exposure (ksec) \\
\tableline
CXO/ACIS$+$HETGS & 2003-05-01T21:41:03 & 48.8 \\
RXTE/PCA 1a & 2003-05-01T17:00:32 & 25.2\\
RXTE/PCA 1b & 2003-05-02T00:02:24 & 0.8 \\
RXTE/PCA 1c & 2003-05-02T00:54:08 & 3.3 \\
\tableline
CXO/ACIS$+$HETGS & 2003-05-28T04:09:21 & 45.5 \\
RXTE/PCA 2a & 2003-05-28T05:28:48 & 1.5 \\
RXTE/PCA 2b & 2003-05-28T06:44:16 & 16.1 \\
RXTE/PCA 2c & 2003-05-28T14:28:32 & 5.6 \\
\tableline
CXO/ACIS$+$HETGS & 2003-06-23T15:56:10 & 50.0 \\
RXTE/PCA 3 & 2003-06-23T17:05:20 & 13.3 \\
\tableline
CXO/ACIS$+$HETGS & 2003-07-30T15:57:58 & 50.2 \\
RXTE/PCA 4a & 2003-07-30T19:48:48 & 0.6 \\
RXTE/PCA 4b & 2003-07-30T21:22:40 & 0.6 \\
RXTE/PCA 4c & 2003-07:31T00:32:48 & 0.6 \\
\tableline
\end{tabular}
\vspace*{\baselineskip}~\\ \end{center}
\tablecomments{The {\it Chandra} and {\it RXTE} observations discussed
in this work are tabulated above.  The start time is given in ``TT''
units.  The net exposure is the exposure after standard {\it Chandra}
and {\it RXTE} filtering has been applied.}
\vspace{-1.0\baselineskip}
\end{footnotesize}
\end{table}

\begin{table}[t]
\caption{Continuum X-ray Spectral Fit Parameters} 
\begin{footnotesize}
\begin{center}
\begin{tabular}{lllllllll}
\tableline
Observation & 1a & 1b & 1c & 2a & 2b	& 2c & 3 & 4 \\          
\tableline

$N_{H}~(10^{22})$ & $2.3$ &	$2.3$	&	$2.3$	&	$2.3$	&	$2.3$	&	$2.3$	&	$2.3$	&	$2.3$\\

$kT$~(keV)	& $1.22(1)$ &	$1.22(1)$ &	$1.21(2)$ &	$1.21(2)$ & $1.23(1)$ &	$1.24(1)$ &	$1.09(1)$ &	$1.009(3)$\\

$K_{MCD}~(10^3)$ & $0.58(1)$ &	$0.88(4)$ &	$0.95(4)$ &	$0.69(4)$ &     $0.62(3)$ &	$0.64(3)$ &	$1.11(3)$ &	$1.01(2)$\\

$\Gamma$	& $2.39(3)$ &	$2.4(1)$ &	$2.6(1)$ &	$2.61(4)$ & $2.63(3)$ &	$2.64(2)$ & 	$2.11(4)$ &	$3.3(3)$\\

$K_{pl}$	& $1.7(1)$ &	$2.0(6)$ &	$3.0(6)$ &	$9(1)$ &        $13.3(6)$ &	$14.8(8)$ &	$0.63(6)$ &	$1.3^{+0.9}_{-0.6}$\\

$F_{3-100}~(10^{-8})$ &	 $1.4(2)$  & $1.9(7)$ &	$1.9(5)$ &	$2.2(4)$ &      $2.6(2)$ &	$2.8(3)$ &	$1.3(2)$ &	$0.6(4)$\\


$f_{hard}$ (3-100) & $0.21$     &	$0.21$ &	$0.21$ &	$0.50$ &     $0.58$ &        $0.57$ &	$0.23$ &	$0.06$\\

$L_{X}$~(3-100)~$(10^{38}~{\rm erg/s})$	&  $1.2(2)$     & 	$1.7(7)$	&	$1.7(4)$	&	$1.9(3)$  &        $2.3(2)$  &	$2.4(2)$ &	$1.1(2)$ &  $0.5(3)$	\\		

$F_{0.5-10}~(10^{-8})$ &  $3.2(5)$        &      $5(2)$ &	$5(1)$ &	$6(1)$ &        $7.0(7)$ &	$7.8(8)$  &	$3.3(4)$ &	$2(1)$\\

$f_{hard}$ (0.5-10) &	 $0.19$             &           $0.15$ &	$0.14$ &	$0.52$ & $0.62$ &	$0.63$ &	$0.09$ &	$0.17$\\

$L_{X}$~(0.5-10)~$(10^{38}~{\rm erg/s})$ &   $2.8(4)$   &	   $4(2)$ &	$4(1)$	&	$5(1)$	&    $6.1(6)$	&	$6.8(7)$	&	$2.9(3)$	& $2(1)$\\

\tableline

$\chi^{2}/\nu$	& $78.8/74$   &	$97.0/74$ &	$98.9/74$ &	$89.5/74$ & $90.2/74$ &	$94.2/74$ &	$117.9/74$ &	$245.1/228$\\

\tableline

Fe~K$\alpha$~(FWHM$=$0~keV)~(eV) & $<4$ &     $<4$ &		$<3$ &		$<7$ & $<9$ &		$<12$ &		$<12$ &		$<6$\\														
Fe~K$\alpha$~(FWHM$=$2~keV)~(eV) & $<6$	 &     $<7$ &		$<7$ &		$<14$ &	$<15$ &		$<20$ &		$<44$ &		$<9$\\

\tableline

\end{tabular}
\vspace*{\baselineskip}~\\ \end{center}
\tablecomments{The results of spectral fits to the {\it RXTE} PCU-2
  and HEXTE-A spectra in the 3--100~keV band are presented above.  The
  observation numbers correspond to those in Table 1.  As observations
  coinciding with the fourth {\it Chandra} spectrum are disk-dominated
  and show little variability, they were fit jointly.  Individual fits
  revealed that $N_{H}$ was consistend with $2.3\times 10^{22}~{\rm
  cm}^{-2}$ for each observation, and was therefore fixed in the fits
  reported here.  The normalization of the power-law component is
  ${\rm photons}~{\rm cm}^{-2}~{\rm s}^{-1}~{\rm keV}^{-1}$ at 1~keV.
  The fluxes quoted above are ``unabsorbed'' fluxes.
  The distance to H~1743$-$322 is unknown; a Galactic
  Center distance of $d=8.5$~kpc was assumed to calculate each
  luminosity value.  All errors are 90\% confidence errors.  The 95\%
  confidence upper-limit on the equivalent width of narrow and broad
  Fe~K$\alpha$ emission lines is quoted at the bottom of the table.}
\vspace{-1.0\baselineskip}
\end{footnotesize}
\end{table}

\clearpage

\centerline{~\psfig{file=f1.ps,angle=-90,width=6.0in}~}
\figcaption[h]{The {\it RXTE} spectrum of H~1743$-$322 from
  observation 2b is shown above, fitted with a multicolor disk
  blackbody plus power-law continuum model (see Table 2).  The spectrum
  is remarkably featureless, with no evidence for disk reflection
  features.  Given the high-frequency (240~Hz) QPOs were
  detected in this data --- indicating the disk is likely very close
  to the assumed black hole --- we might have expected to see strongly
  skewed disk reflection features.}
\medskip

\clearpage

\centerline{~\psfig{file=f2.ps,width=5.0in}~}
\figcaption[h]{Lightcurves from the {\it RXTE} observation 1b are
shown above.  The data have been binned into 32~s bins.  The
variability seen on $few \times 100$s timescales is very similar to
that seen in the {\it Chandra} lightcurves (see Figure 5).}
\medskip

\clearpage

\centerline{~\psfig{file=f3.ps,width=5.0in}~}
\figcaption[h]{The 2--60~keV power density spectrum of {\it RXTE}
observation 1a is shown above.  The Poisson level has been subtracted.
The broad feature at approximately $4\times 10^{-3}$~Hz corresponds to
the variability seen in the lightcurve shown in Figure 2; this feature
can be fit with a Lorentzian.}
\medskip

\clearpage

\begin{table}[t]
\caption{X-ray Absorption Lines in the Spectra of H~1743$-$322}
\begin{footnotesize}
\begin{center}
\begin{tabular}{lllllllllllll}
\tableline
Obs. & \multicolumn{3}{l}{Ion} & Theor. & Meas. & Shift & \multicolumn{2}{c}{FWHM} & Flux & W & N$_{\rm Z}$ & N$_{\rm H}$ \\ 
~ & \multicolumn{3}{l}{~} & (\AA) & (\AA) & (km/s) & ($10^{-3}$\AA) & (km/s) & ($10^{-3}$~ph/cm$^{2}$/s) & (m\AA) & ($10^{17}~{\rm cm}^{-2}$) & ($10^{22}~{\rm cm}^{-2}$) \\

\tableline
 
1 & \multicolumn{3}{l}{Fe XXV} & 1.850 & 1.848(1) & $-320 \pm 160$ & $0.7^{+4.7}_{-0.4}$ & $110^{+740}_{-60}$ & $0.48(8)$ & $1.5(2)$ & $0.62(8)$ & $0.18(2)$\\
1 & \multicolumn{3}{l}{Fe XXVI} & 1.780 & 1.776(1) & $-670 \pm 170$ & $10.6 \pm 2.4$ & $1800 \pm 400$ & $1.4(1)$ & $5.1(5)$ & $4.4(4)$ & $1.3(1)$ \\

\tableline

2 & \multicolumn{3}{l}{Fe XXV} & 1.850 & -- & -- & -- & -- & $<0.5$ & $<1.2$ & $<0.5$ & $<0.2$ \\
2 & \multicolumn{3}{l}{Fe XXVI} & 1.780 & -- & -- & -- & -- & $<0.3$ & $<0.9$ & $<0.8$ & $<0.2$ \\
 
\tableline

3 & \multicolumn{3}{l}{Fe XXV} & 1.850 & 1.851(2) & $+160 \pm 320$ & $2.8^{+6.8}_{-2.8}$ & $450^{+1100}_{-450}$ & $0.25(7)$ & $1.4(4)$ & $0.6(2)$ & $0.18(6)$\\
3 & \multicolumn{3}{l}{Fe XXVI} & 1.780 & 1.778(1) & $-340 \pm 170$ & $6.1^{+4.5}_{-6.1}$ & $1030^{+740}_{-1030}$ & $0.53(8)$ & $3.5(5)$ & $3.0(4)$ & $0.9(1)$ \\

\tableline

4 & \multicolumn{3}{l}{Fe XXV} & 1.850 & 1.850(1) & $0 \pm 160$ & $0.9^{+3.3}_{-0.9}$ & $150^{+550}_{-150}$ & $0.43(4)$ & $4.3(4)$ & $1.8(2)$ & $0.55(6)$\\
4 & \multicolumn{3}{l}{Fe XXVI} & 1.780 & 1.781(1) & $+170 \pm 170$ & $11.3 \pm 3.1$ & $1900 \pm 500$ & $0.54(7)$ & $6.4(8)$ & $5.5(7)$ & $1.7(1)$ \\

\tableline
\end{tabular}
\vspace*{\baselineskip}~\\ \end{center}
\tablecomments{Fit parameters for the He-like Fe~XXV (1s$^{2}$--1s2p)
  and H-like Fe~XXVI (1s-2p) resonance absorption lines detected in
  the combined first-order {\it Chandra}/HEG spectra of H~1743$-$322.
  The errors on all line fit parameters are $1\sigma$ confidence
  errors.  Errors quoted in parentheses are symmetric errors in the
  last digit.  Positive velocity shifts correspond to red-shifts and
  negative velocity shifts correspond to blue-shifts.  Significant
  lines were not detected in observation 2; 95\% confidence
  upper-limits are reported assuming the same line centroid and FWHM
  values as measured in observation 1.  Only the Fe XXVI lines in
  observations 1 and 4 are resolved.  Equivalent neutral hydrogen
  column densities were calculated assuming an Fe abundance of $3.3
  \times 10^{-5}$ relative to hydrogen.  Line wavelengths and
  oscillator strengths are taken from Verner et al. (1996b).}
\vspace{-1.0\baselineskip}
\end{footnotesize}
\end{table}

\begin{table}[t]
\caption{X-ray Absorption Lines in Count-rate-selected Spectra of H~1743$-$322}
\begin{footnotesize}
\begin{center}
\begin{tabular}{lllllllllllll}
\tableline
Obs. & \multicolumn{3}{l}{Ion} & Theor. & Meas. & Shift & \multicolumn{2}{c}{FWHM} & Flux & W & N$_{\rm Z}$ & N$_{\rm H}$ \\ 
~ & \multicolumn{3}{l}{~} & (\AA) & (\AA) & (km/s) & ($10^{-3}$\AA) & (km/s) & ($10^{-3}$~ph/cm$^{2}$/s) & (m\AA) & ($10^{17}~{\rm cm}^{-2}$) & ($10^{22}~{\rm cm}^{-2}$) \\
\tableline
1 ($>$mean) & \multicolumn{3}{l}{Fe XXV} & 1.850 & -- & -- & $<10$ & $<1600$ & $<0.6$ & $<1.8$ & $<0.8$ & $<0.2$ \\
1 ($<$mean) & \multicolumn{3}{l}{Fe XXV} & 1.850 & 1.848(1) & $-320 \pm 160$ & $<10$ & $<1600$ & $0.6(1)$ & $1.9(3)$ & $0.8(1)$ & $0.24(4)$\\
\tableline
1 ($>$mean) & \multicolumn{3}{l}{Fe XXVI} & 1.780 & 1.776(1) & $-670 \pm 170$ & $11.8 \pm 4.7$ & $2000 \pm 700$ & $1.3(2)$ & $4.3(7)$ & $3.7(6)$ & $1.1(2)$ \\
1 ($<$mean) & \multicolumn{3}{l}{Fe XXVI} & 1.780 & 1.776(1) & $-670 \pm 170$ & $10^{+1}_{-2}$ & $1700^{+200}_{-400}$ & $1.4(2)$ & $5(1)$ & $4(1)$ & $1.3(3)$ \\
\tableline
1 (mean$+$5 c/s) & \multicolumn{3}{l}{Fe XXV} & 1.850 & -- & -- & $<10$ & $<1600$ & $<0.8$ & $<0.24$ & $<1$ & $<0.3$ \\
1 (mean$-$5 c/s) & \multicolumn{3}{l}{Fe XXV} & 1.850 & 1.849(1) & $-160 \pm 160$ & $<10$ & $<1600$ & $0.9(1)$ & $2.8(3)$ & $1.2(1)$ & $0.36(4)$\\
\tableline
1 (mean$+$5 c/s) & \multicolumn{3}{l}{Fe XXVI} & 1.780 & 1.775(1) & $-840 \pm 170$ & $11.8 \pm 4.7$ & $2000 \pm 700$ & $1.4(2)$ & $4.6(7)$ & $3.9(6)$ & $1.2(2)$ \\
1 (mean$-$5 c/s) & \multicolumn{3}{l}{Fe XXVI} & 1.780 & 1.777(1) & $-510 \pm 170$ & $14\pm 5$ & $2400\pm 900$ & $1.5(2)$ & $5(1)$ & $4(1)$ & $1.3(3)$ \\
\tableline
3 (norm.) & \multicolumn{3}{l}{Fe XXV} & 1.850 & -- & -- & $<10$ & $<1600$ & $<0.3$ & $<1.5$ & $<0.6$ & $<0.2$\\
3 (dip) & \multicolumn{3}{l}{Fe XXV} & 1.850 & 1.850(1) & $0 \pm 160$ & $<10$ & $<1600$ & $0.7(1)$ & $4.3(6)$ & $1.5(2)$ & $0.45(7)$ \\
\tableline
3 (norm.) & \multicolumn{3}{l}{Fe XXVI} & 1.780 & 1.777(1) & $-510 \pm 170$ & $<10$ & $<1600$ & $0.5(1)$ & $3.1(6)$ & $2.1(5)$ & $0.6(2)$ \\
3 (dip) & \multicolumn{3}{l}{Fe XXVI} & 1.780 & 1.778(1) & $-340 \pm 170$ & $<10$ & $<1600$ & $0.6(1)$ & $4.4(7)$ & $3.8(6)$ & $1.2(2)$ \\
\tableline
\end{tabular}
\vspace*{\baselineskip}~\\ \end{center}
\tablecomments{Fit parameters for the He-like Fe~XXV (1s$^{2}$--1s2p)
  and H-like Fe~XXVI (1s-2p) resonance absorption lines detected in
  the combined first-order count-rate selected {\it Chandra}/HEG
  spectra of H~1743$-$322.  The errors on all line fit parameters are
  $1\sigma$ confidence errors.  Errors quoted in parentheses are
  symmetric errors in the last digit.  Where lines are not detected,
  95\% confidence upper limits are given.  Where measurements are not
  reported, they are fixed to those measured in the time-averaged
  spectra (see Table 3).  Positive velocity shifts
  correspond to red-shifts and negative velocity shifts correspond to
  blue-shifts.  Equivalent neutral hydrogen column densities were
  calculated assuming an Fe abundance of $3.3 \times 10^{-5}$ relative
  to hydrogen.  Line wavelengths and oscillator strengths are taken
  from Verner et al. (1996b).}
\vspace{-1.0\baselineskip}
\end{footnotesize}
\end{table}

\begin{table}
\caption{Absorbing Plasma Parameters}
\begin{footnotesize}
\begin{tabular}{ c r r r r r r r r r}
\tableline
Observation & $n_{rmax}~({\rm cm^{-3}})$ & $r_{min}~({\rm cm})$ & $r_{max}~({\rm cm})$ & $N~({\rm
  cm^{-2}})$ & $\dot{M}_{300}~({\rm g~s^{-1}})$ \\
\tableline
  1         & $1.1 \times 10^{12}$ & $1.2\times 10^{9}$ & $4.0 \times 10^{10}$ & $4.4 \times 10^{22}$ & $1.3 \times 10^{18}$ \\
  2         & $< 1.2 \times 10^{10}$ &  --  & $4.0 \times 10^{11}$ & $< 5.3 \times 10^{21}$ & --- \\
  3         & $2.2 \times 10^{11}$ & $1.5\times 10^{9}$ & $1.0 \times 10^{11}$ & $2.2 \times 10^{22}$ & $1.7 \times 10^{18}$ \\
  4         & $1.8 \times 10^{11}$ &  $3.2\times 10^{7}$ & $1.3 \times 10^{9}$ & $3.7 \times 10^{22}$  & $2.1 \times 10^{14}$ \\

\tableline
\end{tabular}
\vspace*{\baselineskip}~\\
\tablecomments{Parameters for the absorbing plasma model discussed in
  the text are given here.  A single-zone absorbing medium in photoionization
  equilibrium is assumed by the model.  The parameter $\dot{M}_{300}$
  refers to the mass outflow rate assuming a velocity of $300$~km/s.
  The parameter $n_{rmax}$ is the density at $r_{max}$.}
\end{footnotesize}
\end{table}

\clearpage

\centerline{~\psfig{file=f4.ps,width=6.0in}~}
\figcaption[h]{Variable He-like Fe XXV ($\lambda = 1.850$~\AA) and
  H-like Fe XXVI ($\lambda = 1.780$~\AA) absorption lines in the
  combined time-averaged first-order {\it Chandra}/HEG spectra of
  H~1743$-$322 are shown above.  The spectra are numbered in order of
  observation.  The data are plotted in black, 1$\sigma$ error bars
  are plotted in blue, and the model for each spectrum is plotted in
  red.  The continuum models consist of phenomonelogical power-law
  components with Galactic absorption to supply the appropriate
  neutral Fe~K edges from the ISM.  The absorption lines were modeled
  using simple Gaussian components.  There is no significant
  absorption in the spectra from observation 2.}
\medskip

\clearpage

\centerline{~\psfig{file=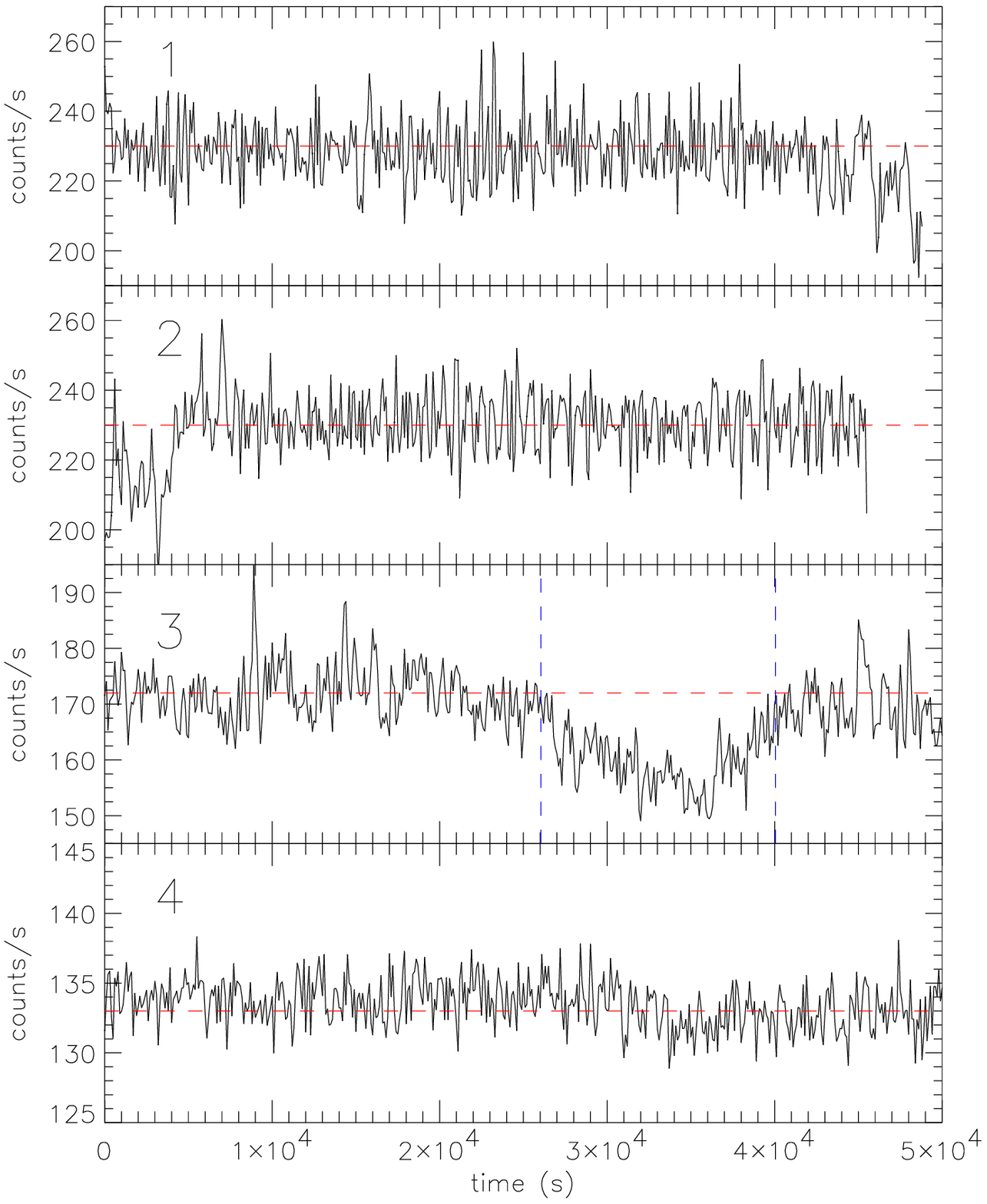,width=6.0in}~}
\figcaption[h]{The 0.5--10.0~keV lightcurves of {\it Chandra}
  observations 1--4 are shown above.  The data were rebinned to have
  bin lengths of 100 seconds.  The mean 1$\sigma$ error in each bin
  above is less than 1.5 counts/s; the extreme variability observed is
  indeed real.  The mean count rate in each observation (excluding the
  dip in observation 3, marked with vertical blue lines) is shown with
  a horizontal red line.  The $\sim$14~ksec dip seen in observation 3
  indicates that H~1743$-$322 is likely viewed at a high inclination;
  weaker dips may be seen at the end of observation 1 and the start of
  observation 2.}
\medskip

\clearpage

\centerline{~\psfig{file=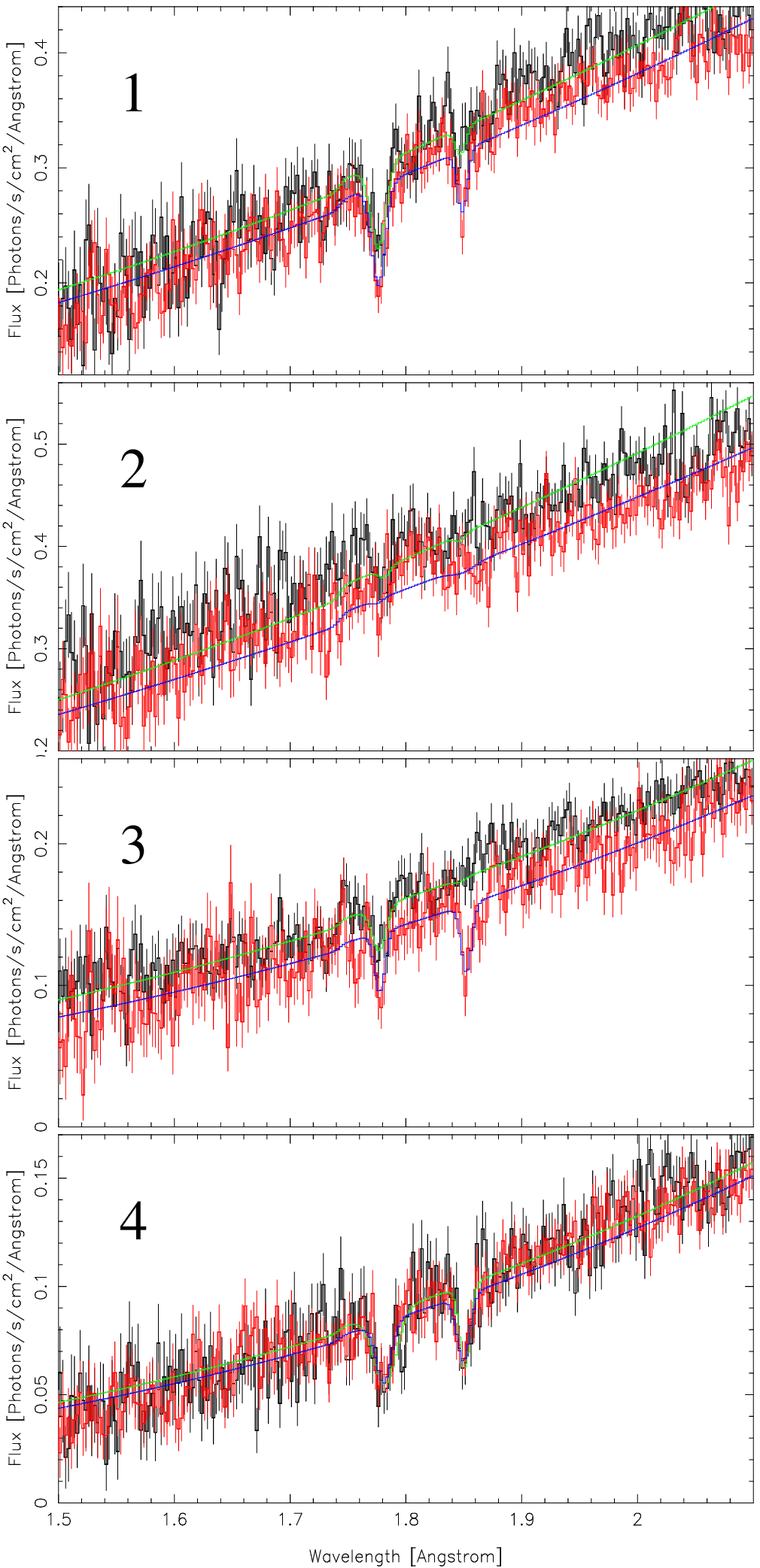,height=8.5in}~}
\figcaption[h]{Count-rate-selected {\it Chandra}/HETGS spectra from
  observations 1--4 are shown above.  In observations 1, 2, and 4, the
black data corresponds to periods above the mean count rate, and the
red data corresponds to periods below the mean count rate.  The
best-fit model for the black data is shown in green, and the best-fit
model for the red data is shown in blue.  In observation 1, there is
evidence that the Fe~XXV ($\lambda = 1.850$~\AA) and Fe~XXVI ($\lambda
= 1.780$~\AA) absorption lines are stronger at low count rates.  The
black data from observation 3 corresponds to the spectrum prior to the
$\sim$14~ksec dip seen in Fig.\ 5, and the red data correspond to the
spectrum within the dip.  The Fe~XXV absorption line is clearly stronger
within the dip.}
\medskip

\clearpage
\centerline{~\psfig{file=f7.ps,width=6.5in}~}
\figcaption[h]{Count-rate selected spectra from Chandra observation 1
  are shown above.  The spectrum in black was taken from time intervals
  with count rates greater than the mean plus 5.0 counts/sec, while
  the spectrum in red was taken from time intervals with count rates
  less than the mean minus 5.0 counts/sec.  These spectra show a more
  distinct change in the depth of the Fe XXV line than those shown in
  panel 1 of Fig.\ 6.}
\medskip

\end{document}